\documentclass{jetpl_arXiv08073896v4}    

\twocolumn

\newcommand{\beq}{\begin{equation}}
\newcommand{\eeq}{\end{equation}}
\newcommand{\beqa}{\begin{eqnarray}}
\newcommand{\eeqa}{\end{eqnarray}}
\newcommand{\bsubeqs}{\begin{subequations}}
\newcommand{\esubeqs}{\end{subequations}}
\newcommand{\half}{{\textstyle \frac{1}{2}}}

\title{$\boldsymbol{f(R)}$ Cosmology from $\boldsymbol{q}$--Theory}

\rtitle{$f(R)$ cosmology from $q$--theory}

\sodtitle{$f(R)$ cosmology from $q$--theory}

\author{F.R. Klinkhamer $^{\#}$\/\thanks{e-mail: frans.klinkhamer@physik.uni-karlsruhe.de}
and G.E. Volovik $^{*+}$\/\thanks{e-mail: volovik@boojum.hut.fi}}

\rauthor{Klinkhamer, Volovik}

\sodauthor{Klinkhamer, Volovik}

\address{$^{\#}$ Institute for Theoretical Physics, University of Karlsruhe (TH),
76128 Karlsruhe, Germany\\
$^*$ Low Temperature Laboratory, Helsinki University of
Technology,
P.O.Box 5100, FIN-02015, HUT, Finland
\\
$^+$ Landau Institute for Theoretical Physics RAS, Kosygina 2,
119334 Moscow, Russia}

\dates{25 July 2008}{*}

\abstract{From a macroscopic  theory of the quantum vacuum
in terms of conserved relativistic charges
(generically denoted by $q^{(a)}$ with label $a$),
we have obtained, in the low-energy limit,
a particular type of $f(R)$ model relevant to cosmology.
The macroscopic quantum-vacuum theory
allows us to distinguish between different
phenomenological $f(R)$ models on physical grounds.}

\begin{document}


\maketitle

\section{Introduction}

The development of high-energy physics and cosmology
over the last years has led to the realization that, most likely,
Einstein's theory of gravity needs to be modified.
At the moment, we are only able to calculate higher-order curvature
corrections to the Einstein action coming
from scalar, spinor, and vector quantum fields
propagating over a fixed classical background
(see, e.g., Refs.~\cite{BirrellDavies1982,CopelandSamiTsujikawa2006}).
The full theory of the modified-Einstein action requires
the underlying microscopic theory  -- what is usually called
\emph{quantum gravity theory}. Since the latter theory is not yet
established, phenomenological models are needed.
Among these are the so-called $f(R)$ models
(see, e.g., Refs.~\cite{CopelandSamiTsujikawa2006,SotiriouFaraoni2008}
for two recent reviews)
which have \emph{ad hoc} powers of the curvature invariants
added to the linear term of the Einstein--Hilbert
action~\cite{Weinberg1972}.

We have proposed another approach to modified Einstein gravity,
which is based on the treatment of the Lorentz-invariant quantum vacuum
as an extended  self-sustained system
characterized by a conserved relativistic
charge  $q$~\cite{KlinkhamerVolovik2008a,KlinkhamerVolovik2008b}. Here,
$q$ is a microscopic variable describing the physics of the
deep (ultraviolet) vacuum, but its thermodynamics and dynamics are
described by macroscopic equations, because $q$ is a conserved quantity.
This quantity $q$ is similar to the particle density in liquids, which
describes a microscopic quantity -- the density of atoms -- but obeys
the macroscopic equations of hydrodynamics, because of particle-number
conservation. Different from known
liquids, the quantum vacuum is Lorentz invariant. The
quantity $q$ must, therefore, be Lorentz invariant.
This treatment has allowed us to discuss both the \emph{thermodynamics}
and the \emph{dynamics} of a Lorentz-invariant quantum vacuum.

The thermodynamic approach~\cite{KlinkhamerVolovik2008a} assumes that
the quantum vacuum  is a stable self-sustained equilibrium state,
which is described by compressibility and
other characteristics of the response to external perturbations.
In this approach, we have found that the vacuum energy
density appears in two forms.

First, there is the microscopic vacuum energy density
which is characterized by an ultraviolet energy scale
$E_\text{UV}$, so that $\epsilon(q)\sim E_\text{UV}^4$.
Most likely, $E_\text{UV}$ corresponds to the standard Planck energy
scale defined in terms of Newton's gravitational constant $G_\text{N}$,
$E_\text{Planck} \equiv \sqrt{\hbar\, c^5/G_\text{N}}
\approx 1.2 \times 10^{28}\:\text{eV}$.
But, here, we prefer to stay as general as possible
and keep $E_\text{UV}$ distinct from $E_\text{Planck}$.

Second, there is the macroscopic vacuum energy density which is
determined by a particular thermodynamic quantity,
\mbox{$\widetilde{\epsilon}_\text{vac}(q) \equiv \epsilon - q\,d\epsilon/dq$,}
and it is this type of energy density which contributes to
the effective gravitational field equations at low energies.
For a self-sustained vacuum in full thermodynamic equilibrium and
in the absence of matter, the effective (coarse-grained) vacuum energy
density $\widetilde{\epsilon}_\text{vac}(q)$ is automatically nullified
(without fine tuning) by the spontaneous
adjustment of the vacuum variable $q$ to its equilibrium value $q_0$,
so that $\widetilde{\epsilon}_\text{vac}(q_0)=0$.
This implies that the effective cosmological constant $\Lambda$
of a perfect quantum vacuum is strictly zero, which is consistent with
the requirement of Lorentz invariance for zero external pressure.

The dynamic approach~\cite{KlinkhamerVolovik2008b} demonstrates how,
in a flat Friedmann--Robertson--Walker universe, the
vacuum energy density $\widetilde{\epsilon}_\text{vac}$
(effective cosmological ``constant'') relaxes from its natural Planck-scale
value at early Planckian times to a naturally small value at late times.

In this Letter, we show that the macroscopic theory of the quantum vacuum,
when applied to cosmology, gives rise to a specific class of $f(R)$ models of
modified gravity.

\section{Gravity with $\boldsymbol{F}$--fields and low-energy matter}
\label{sec:Gravity-with-F-field}

We consider  the general case of several conserved microscopic variables
$q^{(a)}$, for $a=1$, $\dots$ , $n$, and corresponding chemical
potentials $\mu^{(a)}$~\cite{Volovik2004,KlinkhamerVolovik2005}.
As in Refs.~\cite{KlinkhamerVolovik2008a,KlinkhamerVolovik2008b},
each variable $q^{(a)}$ can be represented by a four-form field $F^{(a)}$:
\begin{equation}
(F^{(a)})^2 \equiv -  \frac{1}{24}\,
F^{(a)}_{\mu\nu\rho\sigma}\, F^{(a)\mu\nu\rho\sigma}\,,\quad
F^{(a)}_{\mu\nu\rho\sigma}\equiv
\nabla^{\phantom{(a)}}_{[\mu}\!\!A^{(a)}_{\nu\rho\sigma]}\,.
\label{eq:Fdefinition}
\end{equation}
The action of the four-form fields $F^{(a)}(x)$, the matter field $\phi(x)$,
and the gravitational field $g_{\mu\nu}(x)$ is given by
\begin{eqnarray}
&&
S[A^{(a)}, g,\phi]=
-\int_{\mathbb{R}^4} \,d^4x\, \sqrt{-g}
\nonumber\\
&&
\times \left(
K(F^{(a)})~R   + \epsilon(F^{(a)},\phi)
+ \frac{1}{2}\,\partial_\mu \phi\,\partial^\mu \phi
\right)\,.
\label{eq:actionF}
\end{eqnarray}
For simplicity, low-energy matter is represented by a single real scalar
field $\phi$. The generalized potential $ \epsilon(F^{(a)},\phi)$ includes
self-interactions and interactions between all fields $F^{(a)}$ and $\phi$,
but contains no derivatives of these fields and explicit factors
of the metric field $g_{\mu\nu}$ or its inverse.
The gravitational coupling parameter $K$ is determined by ultraviolet physics
and, therefore, depends on the microscopic vacuum variables $F^{(a)}$.
Here, and in the following, we adopt the conventions of
Ref.~\cite{Weinberg1972}, in particular, those for the Riemann tensor and
the metric signature $(-+++)$.
Natural units with $\hbar=c=1$ are used throughout.

The generalized Maxwell  and Klein--Gordon equations
from action (\ref{eq:actionF}) read
\bsubeqs\label{eq:Maxwell-KleinGordon}
\beqa
\nabla_\mu \left(\sqrt{-g} \;\frac{F^{(a)\mu\nu\rho\sigma}}{F^{(a)}}
\left(  \frac{\partial\epsilon}{\partial F^{(a)}}
       +  R \,\frac{\partial K}{\partial F^{(a)}}
        \right)
                 \right)&=&0\,,
\label{eq:Maxwell}\\
\square\phi -  \frac{\partial\epsilon}{\partial\phi} &=&0\,,
\label{eq:KleinGordon}
\eeqa
\esubeqs
where $\square$ denotes the invariant d'Alembertian operator
and the partial derivatives $\partial/\partial F^{(a)}$ and $\partial/\partial\phi$
stand for pointwise differentiation.
The variation of (\ref{eq:actionF}) over the metric $g_{\mu\nu}$ gives the
generalized Einstein equations:
\begin{eqnarray}
&&
2K
\left( R_{\mu\nu}-\frac{1}{2}\,R\,g_{\mu\nu}\right)
+R\, g_{\mu\nu} \sum_{a=1}^n\, F^{(a)}\,\frac{\partial K}{\partial F^{(a)}}
\nonumber\\
&&
+2
\Big(  \nabla_\mu\nabla_\nu - g_{\mu\nu}\,\square \Big)K
-\widetilde{\epsilon}(F^{(a)},\phi)\, g_{\mu\nu}
+T^\text{M}_{\mu\nu} =0,
\label{eq:EinsteinEquationF}
\end{eqnarray}
with effective potential
\begin{equation}
\widetilde{\epsilon}(F^{(a)},\phi)\equiv \epsilon(F^{(a)},\phi)
-\sum_{b=1}^{n} F^{(b)}\,\frac{\partial\epsilon}{\partial F^{(b)}}  
\label{eq:widetilde-epsilon}
\end{equation}
and scalar-field energy-momentum tensor
\begin{equation}
T^\text{M}_{\mu\nu} =
\partial_\mu \phi\,\partial_\nu \phi
- \frac{1}{2}\,g_{\mu\nu}\,\partial_\rho \phi\,\partial^\rho \phi \,.
\label{eq:widetilde-Tmunu}
\end{equation}

Using $F^{(a)\mu\nu\rho\sigma}$ as given by \eqref{eq:Fdefinition},
the Maxwell equations  (\ref{eq:Maxwell}) can be rewritten in the
following form:
\begin{equation}
\partial_\mu
\left(   \frac{\partial\epsilon}{\partial F^{(a)}}
       + R\,\frac{\partial K}{\partial F^{(a)}}\right)
     =0\,.
\label{eq:Maxwell2}
\end{equation}
The solution of these $4n$
equations is
\begin{equation}
     \frac{\partial\epsilon}{\partial F^{(a)}}
   + R \,\frac{\partial K}{\partial F^{(a)}}
 =\mu^{(a)}  \,,
\label{eq:MaxwellSolution}
\end{equation}
where the $\mu^{(a)}$ are $n$ integration constants.
Eliminating $\partial K/\partial F^{(a)}$
from (\ref{eq:EinsteinEquationF}) and  (\ref{eq:MaxwellSolution}),
one finds for the generalized Einstein equations
\begin{eqnarray}
&&
-2K\Big( R_{\mu\nu}-\half{R}g_{\mu\nu} \Big)
-2\Big(  \nabla_\mu\nabla_\nu - g_{\mu\nu}\, \square\Big)\, K
\nonumber
\\
&&
+\Big(\epsilon-\sum_{a=1}^n \mu^{(a)} F^{(a)} \Big)\, g_{\mu\nu}
=T^\text{M}_{\mu\nu}\,.
\label{eq:EinsteinEquationF2}
\end{eqnarray}

 Equations \eqref{eq:MaxwellSolution} and \eqref{eq:EinsteinEquationF2}
can also be obtained if we use, instead of the original action,
the following effective action:
\begin{eqnarray}
&&
S_\text{eff}[A^{(a)},\mu^{(a)},g,\phi]=
-\int_{\mathbb{R}^4} \,d^4x\, \sqrt{-g}\,
\nonumber
\\
&&
\times \left(
 K\, R  + \epsilon-\sum_{a=1}^n \mu^{(a)} F^{(a)}
+ \frac{1}{2}\,\partial_\mu \phi\,\partial^\mu \phi\right)\,.
\label{eq:ActionMu}
\end{eqnarray}
The  $\mu^{(a)} F^{(a)}$ terms in this action do  not contribute
to the equations of motion \eqref{eq:Maxwell},
because they are total derivatives,
\begin{equation}
\int_{\mathbb{R}^4} \,d^4x\; \sqrt{|g|}\, \mu^{(a)}\, F^{(a)} =
-
\frac{\mu^{(a)}}{24} ~e^{\kappa\lambda\mu\nu}
\int_{\mathbb{R}^4} \,d^4x\; F^{(a)}_{\kappa\lambda\mu\nu} \,.
\label{eq:actionSurface}
\end{equation}
The constant $\mu^{(a)}$ is seen to play the role
of a Lagrange multiplier
related to the conservation of the vacuum ``charge'' $F^{(a)}$.

Instead of the large microscopic energy density $\epsilon(F^{(a)},\phi)$
in the original action \eqref{eq:actionF},
a potentially smaller macroscopic vacuum energy density
enters the effective action \eqref{eq:ActionMu}, namely,
\begin{equation}
 \rho_\text{V}\equiv \epsilon(F^{(a)},\phi)-\sum_{a=1}^n\mu^{(a)} F^{(a)} \,.
\label{eq:ReducedVacEenergy}
\end{equation}
Precisely this macroscopic vacuum energy density gravitates and
determines the cosmological term in the gravitational field equations
\eqref{eq:EinsteinEquationF2}.

\section{Equilibrium vacua and stability conditions}
\label{sec:equilibrium-vacua-stability-conditions}

The main goal of our approach is to describe the thermodynamics
of the equilibrium vacuum~\cite{KlinkhamerVolovik2008a} and to
consider cosmology as the dynamics of relaxation towards an equilibrium
state~\cite{KlinkhamerVolovik2008b,Barcelo2007,Klinkhamer2008}.
That is why, in what follows, we assume that the universe is
close to equilibrium and that all its parameters, including the fields $F^{(a)}$
and the chemical potentials $\mu^{(a)}$, are close to their equilibrium values.
A static  homogeneous equilibrium vacuum, in the absence of thermal matter,
corresponds to a stationary point $\big(F^{(a)}_0, \phi_0\big)$  of
Eqs.~\eqref{eq:KleinGordon}, \eqref{eq:MaxwellSolution},
and \eqref{eq:EinsteinEquationF2} for $R_{\mu\nu}=T^\text{M}_{\mu\nu}=0$:
\begin{equation}
\frac{\partial\epsilon}{\partial\phi}=0
\,,\quad\frac{\partial\epsilon}{\partial F^{(a)}}=\mu^{(a)}
\,,\quad \epsilon -\sum_{a=1}^n\mu^{(a)} F^{(a)}=0\,,
\label{eq:EquilVacuum}
\end{equation}
where the last equation demonstrates that
the vacuum energy \eqref{eq:ReducedVacEenergy} is zero
in an equilibrium vacuum, $\rho_\text{V}|_\text{eq}=0$.

One can see the difference between the conventional matter field $\phi$
and the conserved vacuum fields  $F^{(a)}$,
as only the fields $F^{(a)}$ provide the integration constants $\mu^{(a)}$
which arise dynamically from the solution \eqref{eq:MaxwellSolution}
of the generalized Maxwell equations \eqref{eq:Maxwell2}.
(The field equations of generic matter fields
do not give rise to such integration constants.)
These integration constants $\mu^{(a)}$ play the role of chemical
potentials in thermodynamics and are thermodynamically conjugate to the
density of the conserved quantities $F^{(a)}$.
With appropriate nonzero chemical potentials,
the large vacuum energy $\epsilon(F^{(a)},\phi)$ is reduced to
$\rho_\text{V}=0$ in a homogeneous equilibrium vacuum state
according to \eqref{eq:EquilVacuum}.
Specifically, two large quantities,  $\epsilon(F^{(a)},\phi)$
and $\sum_{a=1}^n\mu^{(a)} F^{(a)}$, each of order $E_\text{UV}^4$,
cancel each other due to
the self-tuning mechanism~\cite{KlinkhamerVolovik2008a}.
This is the main property of a self-sustained vacuum.

For the case of a single vacuum variable $F$,
the chemical potential $\mu$ in equilibrium is completely fixed
by the constraint $\rho_\text{V}=0$.
But, for the case of several variables $F^{(a)}$,
there are $n-1$ degrees of freedom, since the equation $\rho_\text{V}=0$
gives only a single constraint on the $n$ chemical potentials $\mu^{(a)}$.
This allows for the existence of
many different equilibrium vacua and also for
the coexistence of several vacua~\cite{Volovik2004,KlinkhamerVolovik2005}.
This last observation may give microscopic support
to the multiple point principle which
postulates the existence of a number of phases with the same energy density
(see, e.g., Ref.~\cite{FroggattNevzorovNielsen2008} and references therein).

The stationary point $(F^{(a)}_0,\phi_0)$ of the thermodynamic potential
$W\equiv
\epsilon(F^{(a)},\phi)-\sum_a\mu^{(a)} F^{(a)}$
[i.e., the solution of \eqref{eq:EquilVacuum}] should correspond to a minimum,
which can be local or global.
In particular, the vacuum compressibility introduced in
Ref.~\cite{KlinkhamerVolovik2008a} must be positive:
\begin{equation}
\chi_{0} \equiv
\left[\sum_{a,b=1}^{n}
F^{(a)}F^{(b)}\;\frac{\partial^2\epsilon}{\partial F^{(a)}\partial F^{(b)}}\,
\right]_{F^{(a)}=F^{(a)}_0,\;\phi=\phi_0}^{-1}>0\,.
\label{eq:PositiveCompressibility}
\end{equation}
Furthermore, the effective Newton's constant must be positive
in an equilibrium vacuum:
\begin{equation}
G_\text{N}\equiv \frac{1}{16\pi K(F^{(a)}_0)} > 0~,   
\label{eq:PositiveG}
\end{equation}
in order to have a physically consistent description of an attractive gravitational force.
More specifically, negative $K$ gives the wrong sign of the kinetic term for the
graviton (which becomes ghostlike) and the quantum vacuum is unstable.
As $q$--theory is based on a stable self-sustaining vacuum,
having $K>0$  is a necessary condition for stability of the vacuum, together
with the compressibility condition \eqref{eq:PositiveCompressibility}.

\section{$\boldsymbol{\MakeLowercase{f}(R)}$ model
         from $\boldsymbol{\MakeLowercase{q}}$--theory}
\label{sec:f(R)}

Modified-gravity $f(R)$ models appeared already in the 1960s (see, e.g.,
Refs.~\cite{Ruzmaikina1969,Breizman1970}) and were used to construct an
inflationary model of the early universe in the 1980s~\cite{Starobinsky1980}.
More recently, these $f(R)$ models have received renewed attention
as a way to explain the inferred cosmic ``dark energy'' by attributing it
to a modification of Einstein gravity (see, e.g.,
Refs.~\cite{CopelandSamiTsujikawa2006,SotiriouFaraoni2008,Starobinsky2007,KobayashMaeda2008}
and references therein). These models are, in fact, purely phenomenological models,
which, in their simplest form, replace the linear function of the Ricci scalar $R$ from
the Einstein--Hilbert action term by a more general function $f(R)$. This
function $f(R)$ can, in principle, be adjusted to fit the astronomical
observations and to produce a viable cosmological model.

To obtain $f(R)$ from  $q$--theory~\cite{KlinkhamerVolovik2008b}, one can
express $F^{(a)}$ in terms of $R$,  $\phi$, and $\mu^{(a)}$ by use
of the equation system \eqref{eq:MaxwellSolution} and
substitute the resulting functions $F^{(a)}(R)$ into \eqref{eq:EinsteinEquationF2}.
This can be done in a general way
(see Ref.~\cite{SotiriouFaraoni2008});
but, since we consider the relaxation to an equilibrium vacuum, we are only
interested in the simpler situation of a system already close to equilibrium.
In addition, we will restrict ourselves to a single $F$-field
(the generalization to $n$ fields is straightforward) and we also
omit the explicit matter $\phi$-field, keeping only
the general matter energy-momentum tensor $T^\text{M}_{\mu\nu}$.

For a single $F$-field, \eqref{eq:MaxwellSolution} gives
\beqa
\frac{\partial \epsilon(F)}{\partial F}-\mu=
-R\, \frac{\partial K(F)}{\partial F}\,.
\label{eq:EquationFmu-rhoV}
\eeqa
Close to the equilibrium state determined by \eqref{eq:EquilVacuum},
one can expand the microscopic variables as follows:
\beq\label{eq:Fmu-perturbations}
F=F_0+\delta F\,, \quad \mu=\mu_0+\delta\mu.
\eeq
Expressing $\delta F$ in terms of $R$ and $\delta\mu$ and
excluding $\delta F$ from the Einstein equations \eqref{eq:EinsteinEquationF2},
one obtains:
\begin{eqnarray}
&&
-2K\Big( R_{\mu\nu}-\half{R}g_{\mu\nu} \Big)
+2\,\widetilde{\chi}\,\Big(  \nabla_\mu\nabla_\nu\, R - g_{\mu\nu}\, \square\, R
\nonumber
\\
&&
+\frac{1}{4}\,R^2\,g_{\mu\nu}\Big)
-F_0\,\delta\mu\, g_{\mu\nu}
=T^\text{M}_{\mu\nu}\,,
\label{eq:EinsteinEqf(R)}
\end{eqnarray}
in terms of the new dimensionless parameter
\begin{equation}
\widetilde{\chi}\equiv
\chi_{0}\;\Bigg( F \,\frac{\partial K}{\partial F} \Bigg)_{F=F_0}^2\,.
\label{eq:widetildechi}
\end{equation}
In \eqref{eq:EinsteinEqf(R)}, we have omitted the $(\delta\mu)^2$ term
and kept only the leading term containing $\delta\mu$.
Expanding $K$ in the first term of \eqref{eq:EinsteinEqf(R)},
$K=(16\pi G_\text{N})^{-1} +  \delta F\;\partial K/\partial F$,
one obtains the following modified Einstein equations:
\begin{eqnarray}
&&
-\frac{1}{8\pi G_\text{N}}\Big( R_{\mu\nu}-\half{R}g_{\mu\nu} \Big)
+2\,\widetilde{\chi}\Big(  \nabla_\mu\nabla_\nu\, R - g_{\mu\nu}\, \square\, R
\nonumber
\\
&&
-\frac{1}{4}\,R^2\,g_{\mu\nu}+R\,R_{\mu\nu}\Big)
  -F_0\, \delta\mu\,  g_{\mu\nu}
=T^\text{M}_{\mu\nu}\,,
\label{eq:EinsteinEqf(R)2}
\end{eqnarray}
where we have kept only the leading term $F_0\, \delta\mu$ and omitted terms
$R\, \delta\mu$  and $\delta\mu^2$.

The field equations \eqref{eq:EinsteinEqf(R)2} correspond to the following
phenomenological model:
\bsubeqs\label{eq:action+f(R)-derived}
\beqa
S_\text{phenom} &=&
\int_{\mathbb{R}^4} \,d^4x\, \sqrt{-g} \,\left(\frac{1}{16\pi G_\text{N}}\,
\widetilde{f}(R)  + \mathcal{L}^\text{M} \right)\,,\\
\widetilde{f}(R)&=&-R+ 16\pi G_\text{N}\,  \widetilde{\chi}\,  R^2 -2\, \lambda\,,
\label{eq:f(R)-derived}
\eeqa
\esubeqs
with $\mathcal{L}^\text{M}$ the standard matter Lagrange density.
The function $\widetilde{f}(R)$ found belongs to the class of models
$f(R)\sim -R+R^2/(6M^2)$,
where a bare cosmological constant may or may not be added
and $M \equiv (3\,d^2f/dR^2)^{-1/2}$ is the scalaron mass
(see Refs.~\cite{Starobinsky1980,Starobinsky2007} and references therein;
note that our sign convention
for $R$ is opposite to that in Ref.~\cite{Starobinsky2007}). From $q$--theory,
the scalaron mass square is given by
\beq
M^{2}=1\big/\big(\,96 \pi G_\text{N}\, \widetilde{\chi}\,\big)\,,
\label{eq:Msquared-derived}
\eeq
with $\widetilde{\chi}$ defined by \eqref{eq:widetildechi}
for a single ``charge'' $F$ and
by \eqref{eq:widetildechi2} below for multiple ``charges'' $F^{(a)}$.
At this moment, we have two parenthetical remarks.
First, the effective action for \eqref{eq:EinsteinEqf(R)2} does not
contain $R_{\mu\nu}R^{\mu\nu}$ or $R_{\mu\nu\rho\sigma}R^{\mu\nu\rho\sigma}$
terms in addition to the $R^2$ term of \eqref{eq:f(R)-derived}.
Second, a purely phenomenological connection between
generalized-equation-of-state models and $f(R)$ modified-gravity models
has also been noted in App. A of Ref.~\cite{NojiriOdintsov2005}.

In $q$--theory, the  cosmological constant
$\lambda$ in  \eqref{eq:f(R)-derived}  is induced by the deviation
of the chemical potential $\mu$ from its equilibrium value $\mu_0$:
$\lambda\equiv 8\pi G_\text{N}\, \rho_\text{V} = -  8\pi G_\text{N} F_0\delta\mu $.
For the case of $n$ charges $F^{(a)}$, the cosmological constant $\lambda$
is given by
\begin{equation}
\lambda
\equiv  8\pi G_\text{N}\,\rho_\text{V}
=      -8\pi G_\text{N}\,\sum_{a=1}^{n} F_0^{(a)}\delta\mu^{(a)} \,.
\label{eq:lambda}
\end{equation}
For the general case of a quantum vacuum
described by conserved vacuum variables $q^{(a)}$, one has:
\begin{equation}
\Lambda \equiv \frac{\lambda}{8\pi G_\text{N}}
        \equiv \rho_\text{V}
        =      -\sum_{a=1}^{n} q^{(a)}\delta\mu^{(a)} \,.
\label{eq:LambdaGeneral}
\end{equation}
This last equation also follows from the zero-temperature Gibbs--Duhem  
relation~\cite{KlinkhamerVolovik2008a} applied to
the thermodynamic system characterized by several conserved variables:
$\mu^{(a)}$ is the variable thermodynamically conjugate to the conserved
variable $q^{(a)}$.

Note that $f(R)$ phenomenology adds a pair of thermodynamically conjugate
variables,
$R$ and $K$. This follows from, e.g., the thermodynamic potential
$ \epsilon +K\, R -\sum \mu^{(a)} F^{(a)}$    in \eqref{eq:ActionMu}.
The corresponding thermodynamic identities can be used to obtain, for example,
the dimensionless parameter $\widetilde{\chi}$ in
\eqref{eq:f(R)-derived}, which can be interpreted as
$-\partial K/\partial R$ evaluated at $R=0$. One then has, at $R=0$,
\begin{eqnarray}
\widetilde{\chi}&=&
-\frac{\partial K}{\partial R}
=
-\sum_a\frac{\partial K}{\partial F^{(a)}} \,\frac{\partial F^{(a)}}{\partial R}
=\sum_a\frac{\partial K}{\partial F^{(a)}} \,\frac{\partial K}{\partial \mu^{(a)}}
\nonumber\\
&=&
\sum_{a,b}\frac{\partial K}{\partial F^{(a)}} \,
\frac{\partial K}{\partial F^{(b)}}
\frac{\partial F^{(b)}} {\partial \mu^{(a)}}\nonumber\\
&=&
\sum_{a,b}\frac{\partial K}{\partial F^{(a)}}\,
          \frac{\partial K}{\partial F^{(b)}}\,
\left( \frac{\partial^2\epsilon}{\partial F^{(a)}\partial F^{(b)}}\right)^{-1}
\;,
\label{eq:widetildechi2}
\end{eqnarray}
which gives, using definition \eqref{eq:PositiveCompressibility},
precisely \eqref{eq:widetildechi} for the case of a single charge $F$.
However, this thermodynamic description
is only applicable to $f(R)$ phenomenology,
since it does not take into account non-$f(R)$ terms
in the action such as $R_{\mu\nu}R^{\mu\nu}$  and
$R_{\mu\nu\rho\sigma}R^{\mu\nu\rho\sigma}$ which appear  due to quantum
corrections~\cite{BirrellDavies1982,CopelandSamiTsujikawa2006}.

For $q$--theory of the type considered in \eqref{eq:actionF},
there are two stability conditions, namely,
condition \eqref{eq:PositiveG} for the effective Newton's
constant $G_\text{N}$ and condition \eqref{eq:PositiveCompressibility}
for the zero-temperature vacuum compressibility $\chi_0$.
These two conditions of $q$--theory correspond, respectively, to the following stability
conditions of $f(R)$ models (see, e.g., Eq.~(7) of Ref.~\cite{Starobinsky2007}):
\begin{equation}
-f'(R)\,\Big|_{R=0}>0\,,\quad f''(R)\,\Big|_{R=0}>0 \,,
\label{eq:f(R)stability}
\end{equation}
where the prime denotes differentiation with respect to $R$ and the extra
minus sign traces back to our conventions,
as mentioned already a few lines under \eqref{eq:f(R)-derived}.
These conditions ensure the stability of the
empty flat universe, which is a basic assumption for $q$--theory as it
concerns the thermodynamics of the equilibrium vacuum.
In $f(R)$ models, the second condition of \eqref{eq:f(R)stability},
i.e., the positive mass squared
of the scalaron, $M^2>0$, implies the stability of the Minkowski vacuum,
whereas for the negative  mass squared, $M^2<0$,  the Minkowski vacuum
experiences the scalaron instability~\cite{Starobinsky2007}.

For the case of $M^2>0$, the relaxation of the universe
to equilibrium is accompanied by oscillations of $R$
with frequency $M$ (see Ref.~\cite{Starobinsky2007} for the $f(R)$ model
and Ref.~\cite{KlinkhamerVolovik2008b} for the $q$--theory
which also has $\delta F$ oscillations).
In particular, the vacuum energy density in $q$--theory   
has been found to relax as follows~\cite{KlinkhamerVolovik2008b}:
  \begin{equation}
\rho_\text{V} \sim \frac{M^2}{t^2}~ \sin^2 Mt \,,
\label{eq:ocillations}
\end{equation}
for cosmic time $t \gg 1/M$.
While the $q$ field itself has an equation of state parameter $w=-1$,
corresponding to vacuum energy density and a cosmological constant,
the decay of the vacuum energy density in \eqref{eq:ocillations} simulates
the evolution of cold dark matter with  $w=0$.
The effective equation of state $w=0$ is induced
by the interaction of the $q$ field with gravity~\cite{KlinkhamerVolovik2008b}.

\vspace*{0mm}
\section{Comparison with phenomenological $\boldsymbol{\MakeLowercase{f}(R)}$ models}
\label{sec:Comparison}

As explained above and, more briefly, in Ref.~\cite{KlinkhamerVolovik2008b},
$F$ theory (or, more generally, $q$--theory)
may give a microscopic justification for the phenomenological $f(R)$ models
used in theoretical cosmology and may allow for a choice between different
classes of model functions $f(R)$ based on fundamental physics.
Close to equilibrium, the  effective $f(R)$ model emerging
from the $q$--theory of quantum vacuum
belongs to the class of $f(R)\sim -R+R^2/M^2$ models.
This rules out so-called $1/R$ models,
i.e., models with $f(R)\sim -R+M^4/R$ (see, e.g.,
Refs.~\cite{CopelandSamiTsujikawa2006,SotiriouFaraoni2008,KobayashMaeda2008}
and references therein).

Here, we have assumed that the physics of the vacuum variable in
$q$--theory~\cite{KlinkhamerVolovik2008a,KlinkhamerVolovik2008b}
is determined by a unique microscopic energy scale
$E_\text{UV}$. This implies the following orders of magnitude:
\bsubeqs
\beqa
|F_0^{(a)}|\sim  |\mu_0^{(a)}| &\sim&E_\text{UV}^2\,,\\
|F_0^{(a)}\partial K/\partial F_0^{(a)}|\sim G_\text{N}^{-1}&\sim& E_\text{UV}^2\,,\\
|\epsilon(F_0^{(a)}, \phi_0)| &\sim& E_\text{UV}^4\,,\\   
\chi_{0} &\sim&  E_\text{UV}^{-4}\,.
\eeqa
\esubeqs
As a result, the dimensionless parameter $\widetilde{\chi}$
in \eqref{eq:widetildechi} can be expected to be of order unity.
If the assumption of a single fundamental energy scale holds true, one then
obtains that the mass parameter \eqref{eq:Msquared-derived} of the induced
$f(R)=-R+R^2/(6 M^2)$
model is of the order of the ultraviolet scale, $M\sim E_\text{UV}$.
In  this case, the $R^2$ correction is of the order of the quantum $R^2$
corrections to the Einstein
action~\cite{BirrellDavies1982,CopelandSamiTsujikawa2006}.
Moreover, as $q$--theory starts from the assumption that
the vacuum is a stable self-sustained medium, one has $M^2>0$,
as follows explicitly from \eqref{eq:PositiveCompressibility},
\eqref{eq:widetildechi}, and \eqref{eq:f(R)-derived}.

All this agrees with the simplest $f(R)\sim -R+R^2/M^2$ model
having a Planck-scale mass $M$~\cite{Starobinsky1980},
but disagrees with certain other models which use
more complicated phenomenological functions $f(R)$.
For the particular
models
suggested in Ref.~\cite{Starobinsky2007},
the functions $f(R)$ expanded around $R=0$
have negative $M^2$ (making Minkowski spacetime unstable)
and $-M^2$ of the order of $\lambda_\text{\,now}$, where
$\lambda_\text{\,now} = 8\pi G_\text{N}\,\rho_\text{V,\,now}
\approx (10^{-33}\:\text{eV})^2$
is the present (positive) cosmological constant
as determined by observational cosmology~\cite{Komatsu-etal2008}.
These cosmologically desirable $f(R)$ models can, in principle, be
obtained by using special choices of $\epsilon(F^{(a)},\phi)$   
and $K(F^{(a)})$. But such choices require a careful tuning
of the parameters, which is unnatural from our point-of-view
on the properties of the self-sustained quantum vacuum.

\vspace*{0mm}
\section{Conclusion}
\label{sec:Conclusion}

In $q$--theory~\cite{KlinkhamerVolovik2008a,KlinkhamerVolovik2008b}
with chemical potentials $\mu^{(a)}$ at
their equilibrium values  $\mu^{(a)}_0$, the
vacuum energy density $\rho_\text{V}$
has been found to relax, according to \eqref{eq:ocillations},
from its  natural Planck-scale value
at early times when the system is far from equilibrium
to a naturally small value at late times when the system is close to equilibrium.
This solves the main cosmological problem:
the present cosmological constant is small compared to Planck-scale values
simply because the universe happens to be old compared to Planck-scale times.
The remaining problem is to understand why the cosmological constant does
not completely relax to zero as $t\rightarrow \infty$ or,
in other words, to determine the origin of the small residual part of the vacuum energy
which remains (almost) constant during the present epoch.

The suggested $f(R)$ model in which the
cosmological constant appears just at the latest stage
\cite{Starobinsky2007} does not follow from
the macroscopic quantum-vacuum approach.
That is why one needs to
find another explanation for the observed value of $\Lambda$.
In order to produce a nonzero $\Lambda$,
one must find, according to \eqref{eq:lambda}--\eqref{eq:LambdaGeneral},
processes which shift the chemical potentials $\mu^{(a)}$ away
from the equilibrium mean-field values $\mu^{(a)}_0$.

There has been the suggestion to relate the present value of $\Lambda$
to quantum fluctuations of the vacuum energy, for the case that
the vacuum energy itself is nullified (see, e.g.,  Ref.~\cite{Padmanabhan2008}).
This approach could be helpful in $q$--theory, where the vacuum energy is
necessarily zero in full equilibrium.
The observed nonzero value of $\Lambda$ would then correspond
to quantum or thermodynamic fluctuations of the chemical potentials
$\mu^{(a)}$ compared to the equilibrium mean-field values
$\mu^{(a)}_0$. Still, there could be other processes which are able to
shift $\mu^{(a)}$ somewhat away from the mean-field values $\mu^{(a)}_0$,
one example being the backreaction of matter
and another example being matter production by the rapid oscillations
which accompany the relaxation of the vacuum energy \eqref{eq:ocillations}.
It remains to be seen which of these processes, if any,              
is the dominant one for the inferred small but nonzero value of the
vacuum energy density from observational cosmology.

\vspace*{0mm}
\section*{\hspace*{-4.5mm}ACKNOWLEDGMENTS}
\noindent
It is a pleasure to thank A.A. Starobinsky for helpful
discussions. GEV is supported in part by the Russian Foundation for
Basic Research (grant 06--02--16002--a) and the Khalatnikov--Starobinsky
leading scientific school (grant 4899.2008.2)


\end{document}